\documentclass[prl,twocolumn,lengthcheck]{revtex4-1}

\usepackage{amsmath}
\usepackage{amssymb}
\usepackage{graphicx}
\usepackage{hyperref}
\usepackage{color}
\usepackage{braket}

\newcommand{\tr}{\operatorname{tr}}

\begin{document}

\title{Stochastic Integral Representation for the Dynamics of Disordered Systems}

\author{Ivana Kure\v{c}i\'c$^{1}$}
\author{Tobias J.\ Osborne$^{2}$}
\affiliation{$^{1}$Max Planck Institute of Quantum Optics, Hans-Kopfermann-Str. 1, 85748 Garching, Germany \\
$^{2}$Institut f\"ur Theoretische Physik, Leibniz Universit\"at Hannover, Appelstr.\ 2, 30167 Hannover, Germany}

\begin{abstract}
The dynamics of interacting quantum systems in the presence of disorder is studied and an exact representation for disorder-averaged quantities via It\^o stochastic calculus is obtained. The stochastic integral representation affords many advantages, including amenability to analytic approximation, applicability to interacting systems, and compatibility with existing tensor network methods. The integral may be expanded to produce a series of approximations, the first of which already includes all diffusive corrections and, further, is manifestly completely positive. The addition of fluctuations leads to a convergent series of systematic corrections.  As examples, expressions for the density of states, spectral form factor, and out-of-time-order correlators for the Anderson model are obtained.
\end{abstract}

\maketitle
Since Anderson's disovery of the \emph{localisation phenomenon} \cite{anderson_absence_1958}, quantum systems in the presence of disorder have attracted considerable interest. The localisation phenomenon is a result of quantum coherent effects, thus presenting many challenges and exhibiting rich physics. During the past 60 years there has been excellent progress in the study of disordered systems in the single-particle regime, culminating in a solid understanding of these physical phenomena (see, e.g., the recent review \cite{lagendijkFiftyYearsAnderson2009}). These advances are the result of several powerful techniques which have been developed to quantitatively study disordered systems, including numerical methods \cite{schenkLargeScaleDiagonalizationTechniques2008}, the renormalisation group \cite{abrahams_scaling_1979}, and field-theoretic methods \cite{wegnerMobilityEdgeProblem1979} such as the replica trick \cite{altshulerQuantumPhenomenaMesoscopic,lernerStronglyCorrelatedFermions2002}, the supersymmetric method \cite{efetov_supersymmetry_1997,zirnbauerSupersymmetrySystemsUnitary1996}, and the Keldysh formalism \cite{kamenev_keldysh_2009,kamenev_electron-electron_1999,kamenevFieldTheoryNonEquilibrium2011}. 

With the recent discovery of the intriguing phenomenon of \emph{many body localisation} (MBL) \cite{baskoMetalInsulatorTransition2006} (see also the recent reviews \cite{alet_many-body_2018,nandkishoreManyBodyLocalizationThermalization2015,altmanUniversalDynamicsRenormalization2015,abaninRecentProgressManybody2017}), renewed interest in disordered systems has emerged. As a collective many particle effect, MBL is proportionally more challenging to study than its single-particle counterpart. The most progress in this field has come from perturbation theory and direct numerical simulation via Monte Carlo sampling. In general, these techniques are limited either to weak disorder or finite system sizes (of the order of 20 spins), and few samples. Exploiting the most powerful field-theoretic approaches to study the strongly interacting quantum spin systems exhibiting MBL appears to be a difficult task (for some recent progress, see \cite{liaoResponseTheoryErgodic2017}).

Some intriguing new directions --- emerging from holographic arguments in high energy physics \cite{breakthroughAlexeiKitaev2015} --- are also being explored in the study of the complex dynamics of quantum systems such as the Anderson model \cite{swingleSlowScramblingDisordered2017}. Here a central role is played by \emph{out-of-time-order correlation functions} (OTOCs) \cite{larkinQuasiclassicalMethodTheory1969} as signatures of quantum chaos. The behaviour of OTOCs for complex disordered systems is a central goal in the study of new phases of disordered matter.   

Motivated by the twin challenges of MBL and calculating OTOCs for disordered systems, in this Letter we introduce a stochastic integral representation for the disorder-averaged propagator of an arbitrary quantum system. The derivation of this representation is reminiscent, in parts, of field-theoretic approaches to disordered systems and also separately to recent calculations \cite{kosManyBodyQuantumChaos2018,bertiniExactSpectralForm2018} of Prosen and coworkers. However, there are crucial differences. For example, there is no mapping to the nonlinear sigma model. Also, the representation is provably exact and not an approximation. We exploit the integral to produce a series of approximations, the first of which already includes all diffusive corrections and, further, is manifestly the result of a completely positive evolution, thus conserving probabilities. The addition of fluctuations leads to a convergent series of systematic corrections. Using this expansion we compute the density of states and out-of-time-order correlations for the Anderson model.

\paragraph{Disordered quantum systems, Brownian motions, and stochastic calculus.---}\hspace{-1em}Here we introduce the systems under consideration and give a brief notational summary of stochastic calculus. We consider quantum systems with a Hamiltonian of the form
\begin{equation}\label{eq:disham}
	H(\mathbf{x}) = H_0 + \sum_{j=1}^m x_j D_j,
\end{equation}
where $H_0$ is a fixed Hamiltonian (typically a kinetic energy term), $D_j$ represent disordered terms (e.g., a local magnetic field or potential energy terms), and $x_j$ are random variables drawn from the Gaussian distribution \footnote{The restriction to a Gaussian-distributed disorder is not fundamental and the techniques developed here are easily generalised to any \emph{divisible} disorder distribution.} with the probability density function
\begin{equation*}
	 g_\gamma(x) \equiv \frac{e^{-\frac{x^2}{2\gamma^2}}}{\sqrt{2\pi}\gamma},
\end{equation*}
where $\gamma$ is its standard deviation. This class of models is sufficiently general to describe a diverse variety of models from single impurity models, the Anderson model in arbitrary dimensions, as well as MBL systems. We illustrate the results in terms of the one-dimensional Anderson tight-binding model with periodic boundary conditions, where
\begin{equation}
	H_0 \equiv 2\mathbb{I} -\sum_{j=1}^N  |j+1\rangle \langle j| + |j\rangle \langle j+1|,
\end{equation}
and $D_j \equiv |j\rangle\langle j|$. We study disorder-averaged dynamical quantities, such as the disorder-averaged propagator:
\begin{equation}
	S(t) = \mathbb{E}_{\mathbf{x}}[ e^{itH}],
\end{equation}
whose Fourier transform yields the density of states (DOS), the
density operator:
\begin{equation}
	\rho(t) = \mathbb{E}_{\mathbf{x}}[ e^{-itH}|\psi_0\rangle\langle \psi_0|e^{itH}],
\end{equation}
spectral form factors:
\begin{equation}\label{eq:sff}
	\mathbb{E}_{\mathbf{x}}[|\tr(e^{(-\beta-it)H})|^{2k}],
\end{equation}
and OTOCs:
\begin{equation}\label{eq:otoc}
	\mathbb{E}_{\mathbf{x}}[\langle A(0)B(t)C(0)D(t)\rangle].
\end{equation}
These quantities are intimately interrelated (see, e.g., \cite{cotlerChaosComplexityRandom2017a}), and can all be calculated in terms of $S(t)$ by taking tensor copies of the Hamiltonian \footnote{For example, we can calculate $\rho(t)$ from $S_2(t) = \mathbb{E}_{\mathbf{x}}[ e^{it(H\otimes \mathbb{I} - \mathbb{I}\otimes H^T)}]$.}.

We make use of stochastic calculus, but it is not at all necessary to be familiar with this formalism, as all derivations can be understood with little more than a passing familiarity with Gaussian integrals and a tolerance for lengthy derivations using discretisations \footnote{The elementary derivation via Gaussian integrals of the main representation is carried out in the supplementary material.}. At this stage it is sufficient to comment that a \emph{Brownian motion} or \emph{Wiener process} $W_t$ is characterised by the following four properties: (1) $W_0= 0$; (2) $W_t$ is (almost surely) continuous; (3) $W_t$ has independent increments; and (4) $W_t-W_s$ is distributed according to the normal distribution with a mean of zero and variance $t-s$ for $0\le s\le t$. We will also encounter stochastic differential equations (SDE) of the form $dy = f\, ds + g\, dW$. At this stage, it is sufficient to regard these as equations that, upon discretisation, define new random variables $y$ in terms of $W$. For further details on stochastic calculus see, e.g., \cite{oksendalStochasticDifferentialEquations2003,gardinerHandbookStochasticMethods1997}.

\paragraph{A stochastic integral representation.---}\hspace{-1em}
In this section we summarise the salient features of the derivation of our integral representation, with an emphasis on the physical foundations of the argument. As explained previously, for Hamiltonians of the form (\ref{eq:disham}), it is sufficient to restrict our attention to the study of the quantity
\begin{equation}
	S(t) \equiv \mathbb{E}_{\mathbf{x}}[e^{itH(\mathbf{x})}].
\end{equation} 
For simplicity, we focus on Hamiltonians of the form $H(x) = A+xB$, where $x$ is a single Gaussian-distributed random variable, and $A$ and $B$ are $N\times N$ matrices. (The extension to Hamiltonians of the form from Eq.~(\ref{eq:disham}) is entirely straightforward and requires no additional techniques.) We derive our representation in four steps. The first step is to break the propagator $e^{itH}$ into $n$ small pieces $\left(e^{\frac{it}{n}H}\right)^n$, introduce $n$ independent Gaussian-distributed variables $x_j$, $j=1, 2, \ldots, n$, and enforce equality via delta functions:
\begin{equation}
  S(t) = \frac{1}{\sqrt{2\pi}\gamma}\int  e^{-\frac{\|\mathbf{x}\|^2}{2n\gamma^2}}\boldsymbol{\delta}(f(\mathbf{x})) \prod_{j=1}^n e^{\frac{it}{n}(A+x_jB)}\,d\mathbf{x},
\end{equation}
where the product is taken from right to left and $\boldsymbol{\delta}(f(\mathbf{x})) = \delta(x_2-x_1)\cdots \delta(x_n-x_{n-1})$.
The second step is to use the identity
\begin{equation}
	\delta(x) \equiv \frac{1}{2\pi}\int_{-\infty}^\infty e^{ikx}\, dk,
\end{equation}
and to carry out the integral over the $x_j$ variables. This leaves an integral over $k_1, k_2, \ldots, k_{n-1}$:
\begin{equation}
  S(t) = \frac{1}{(2\pi)^{n-\frac12}\gamma}\int   e^{-\frac{n\gamma^2}{2}\sum_{j=1}^{n}(k_{j-1}-k_j)^2} \widetilde{F}(\mathbf{k})\, d\mathbf{k},
\end{equation}
where we've introduced two additional $k$ variables, $k_0 = k_{n} = 0$, as well as the function
\begin{equation}
	\widetilde{F}(\mathbf{k}) \equiv \prod_{j=1}^n \widetilde{F}_j(k_{j-1}-k_j),
\end{equation}
where
\begin{equation}
	\widetilde{F}_j(y) = \int_{-\infty}^\infty e^{-\left(\frac{1}{\sqrt{2n}\gamma}x_j-i\sqrt{\frac{n}{2}}\gamma y\right)^2} e^{\frac{it}{n}(A+x_jB)}\, dx_j.
\end{equation}
The third step is to expand the exponents of the operator exponentials in $\widetilde{F}$s, collect terms to $O(1/n^2)$, and then carry out the integrals over $x_j$:
\begin{equation}
	\int \prod_{j=1}^n F(k_{j-1}-k_j)\, d\mu,
\end{equation}
where 
\begin{equation}
	F(y) \equiv \mathbb{I} + \frac{it}{n}(A+i\gamma^2n y B ) - \frac{t^2}{2n^2}(n\gamma^2 - n^2\gamma^4y^2 )B^2
\end{equation}
and
\begin{equation}
	d\mu \equiv \sqrt{\frac{n(n\gamma^2)^{n-1}}{(2\pi)^{n-1}}} e^{-\frac{n\gamma^2}{2}\mathbf{k}^T \mathbf{M} \mathbf{k}} \, dk_1\cdots dk_{n-1},
\end{equation}
with
\begin{equation}
	\mathbf{M} = \begin{pmatrix}
		2  & -1 & 0  & 0   &\cdots & 0  \\
		-1 & 2  & -1 & 0   &\cdots & 0  \\
		0  & -1 & 2  & -1  &       & 0    \\
		\vdots   &    &    &     &\ddots & \vdots      \\
		0  &  \cdots  &    &	0   &    -1   & 2  
	\end{pmatrix} =  2\mathbb{I} -P_{n-1}.
\end{equation}
In physical terms, $\mu$ is the (discretisation of the) equilibrium probability distribution for a free particle with the Hamiltonian $H = -\nabla^2$, which is diffusively moving on the interval $[0,1]$ with vanishing Dirichlet boundary conditions.

By employing the approximation
\begin{equation}
	F(y) \approx e^{\frac{it}{n}\left(A+in\gamma^2(k_{j-1}-k_j)B\right) - \frac{t^2}{2n}\gamma^2B^2},
\end{equation}
valid to $O(1/n^2)$, we have already arrived at an approximation of great utility in tensor-network simulations (this will be the subject of a forthcoming paper):
\begin{equation}
	S(t) \approx \int \prod_{j=1}^n e^{\frac{it}{n}A - t\gamma^2(k_{j-1}-k_j)B - \frac{t^2}{2n}\gamma^2B^2}\, d\mu + O(1/\sqrt{n}).
\end{equation}
The final step is to recognise the integral measure in the limit of $n\rightarrow \infty$ as the path measure for the \emph{Brownian bridge} \cite{oksendalStochasticDifferentialEquations2003}, which is a continuous-time stochastic process with the same conditional probability distribution as the Wiener process, but subject to the condition that at $s=0$ and $s=1$ it is pinned to $0$, i.e., $B_1 = 0$. The Brownian bridge is defined by
\begin{equation}
	z_s = W_s - sW_{s=1}.
\end{equation}
Note that the increments of the Brownian bridge are not independent. This allows us to take the continuum limit:
\begin{equation}\label{eq:bbrep}
	S(t) = \int  \mathcal{T}e^{\int_0^1 K\,ds + \gamma t\int_0^1 B \, dz}\, d\mu,
\end{equation}
where $\mathcal{T}$ is the time-ordering operation, 
\begin{equation}
	K = it A - \frac{\gamma^2t^2}{2}B^2,
\end{equation}
and the increment $dz$ obeys the stochastic differential equation:
\begin{equation}
	dz = -\frac{z}{1-s}\,ds + dW.
\end{equation}
It is important to note that Eq.~(\ref{eq:bbrep}) is an \emph{equality} --- this formula is not an approximation. In this way we have obtained a representation of the operator $S$ via the operator SDE:
\begin{equation}
	dS = it A S\, ds + \gamma tB S\, dz.
\end{equation}
This representation may be subjected to a variety of solution and approximation techniques, from direct sampling, moment expansions, and the Dyson series. These will all be the subject of future studies.

By following the derivation described above we can immediately write down the stochastic integral representation for Hamiltonians $H$ of the form Eq.~(\ref{eq:disham}):
\begin{equation}
	S(t) = \int \mathcal{T}e^{\int_0^1 K \, ds + \gamma t\sum_{j=1}^m\int_0^1  D_j \,dz_j} \,  d\mu(z) ,
\end{equation}
where 
\begin{equation}
	K = it H_0  -\frac{\gamma^2 t^2}{2} \sum_{j=1}^m D_j^2,
\end{equation}
with
\begin{equation}
	dz_j = -\frac{z_j}{1-s}\,ds + dW_j,
\end{equation}
and $W_j$ are $m$ independent Brownian motions.

An alternative derivation of our representation may be found by employing the Lie-Trotter formula $e^{A+B} \approx e^Ae^B + O(\|[A,B]\|^2)$ and an operator Hubbard-Stratonovich transformation.

\paragraph{The stochastic Dyson series: diffusions and fluctuations.---}\hspace{-1em}In this section we describe a Dyson series procedure to develop the integral in a power series in the disorder parameter $\gamma$. This series has several extremely desirable features with the following physical interpretations. The first term in the series already explicitly incorporates disorder corrections and describes a completely positive evolution with diffusive behaviour. The subsequent terms incorporate quantum \emph{fluctuation} corrections around the diffusions. This solution exhibits excellent large $t$ behaviour (in contrast to a direct Dyson series approximation of the propagator followed by a disorder average).

We focus, again for simplicity, on the simplified case $H = A+xB$, and make an expansion of the integral Eq.~(\ref{eq:bbrep}) for $S(t)$ in powers of the stochastic term $\gamma t \int_0^1 B\, dz$ in the exponent. We do this by first defining 
$B(s) \equiv e^{s K} B e^{-s K}$ and writing
\begin{equation}
	e^{-K}S(t) = \int \mathcal{T}e^{\gamma t\int_0^1 B(s) \,dz} \,  d\mu(z).
\end{equation}
Expanding the exponential with a standard Dyson series leads to
\begin{multline}
	e^{-K}S(t) = \mathbb{E}\bigg[\mathbb{I} + \gamma t \int_0^1 B(s) \,dz + \\ \frac{\gamma^2 t^2}{2} \int_0^1\int_0^1 \mathcal{T}[B(s_1)B(s_2)] \,dz_1dz_2 + \cdots \bigg].
\end{multline}
The next step is to employ the covariance of the Brownian bridge, $\mathbb{E}[z(s)z(t)] = \min\{s(1-t), t(1-s)\} = C_{st}$, from which we derive $\mathbb{E}[dz(s)dz(t)] = (\delta(s-t)-1) dsdt$,
so that, to $O(\gamma^2)$, we have
\begin{multline}\label{eq:dyson}
	e^{-K}S(t) = \mathbb{I} - {\gamma^2 t^2} \int_0^1 ds_1\,\int_0^{s_1}ds_2\, B(s_2)B(s_1)   + \\ \frac{\gamma^2 t^2}{2} \int_0^1 B(s)^2 \,ds.
\end{multline}
(We can compute higher-order terms using the classical Wick's theorem, e.g., $\mathbb{E}[z(s_1)z(s_2)z(s_3)z(s_4)] = C_{s_1s_2}C_{s_3s_4} + C_{s_1s_3}C_{s_2s_4} + C_{s_1s_4}C_{s_2s_3}$. This will be the subject of a future paper.)

The expansion Eq.~(\ref{eq:dyson}) admits a very pleasing physical interpretation: the $O(1)$ term $S_1(t) = e^{itA-\frac{\gamma^2 t^2}{2}B^2}$ already explicitly incorporates the effects of disorder in the form of \emph{diffusive corrections} $e^{-\frac{\gamma^2 t^2}{2}B^2}$. When we apply this technique to the disorder-averaged density operator $\rho(t)$, it leads to the expression
\begin{equation}
	\rho(t) \approx e^{\mathcal{L}}[\rho(0)],
\end{equation}
where
\begin{equation}
	\mathcal{L}(X) \equiv it [A, X] - \frac{\gamma^2 t^2}{2} \{B,X\} + \gamma^2 t^2BX B
\end{equation}
is a generator of \emph{Lindblad form}, meaning that the evolution $e^\mathcal{L}$ is completely positive, hence physical. 

Because the diffusive solution generically supplies an exponential suppression in $t$, we see that all the subsequence fluctuation corrections are exponentially suppressed. Further, one can argue that the resulting series in $O(\gamma)$ is actually convergent, in contrast to some field-theoretic approaches.

\begin{figure}
	\includegraphics{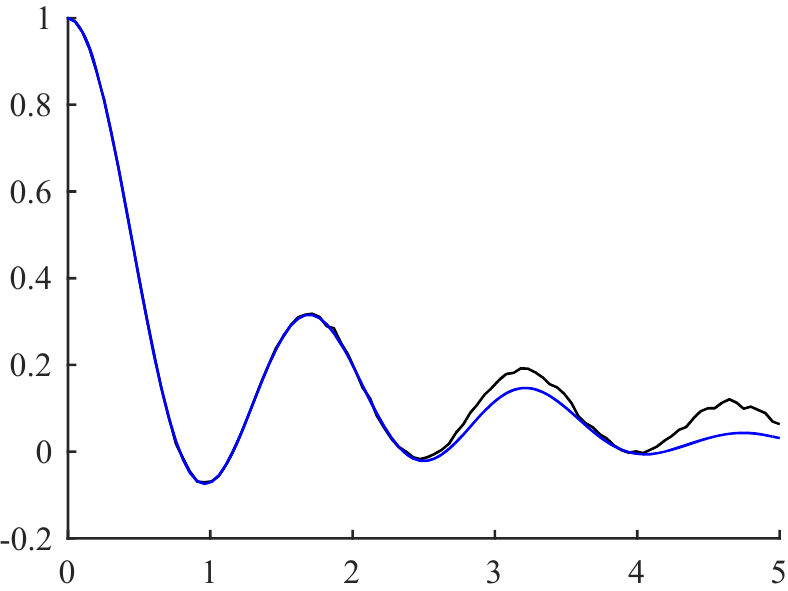}
	\caption{The quantity $X(t)\equiv \frac{1}{N}\mathbb{E}[\tr(e^{itH})]$ for the Anderson model on 30 sites (the $x$ axis is time in units where $\hbar = 1$). (It turns out that the results for $30$ sites are already indistinguishable from the $N= \infty$ limit.) Shown black is the result of numerical sampling with $100$ samples. Shown blue is $X(t) = X_0(t)+X_2(t)$, the second order result calculated via the stochastic Dyson series Eq.~(\ref{eq:2ndorderX}).}\label{fig:x}
\end{figure}

\paragraph{Application 1: the density of states for the Anderson model.---}\hspace{-1em}Here we detail the calculations for the Anderson model. We focus on the quantity
\begin{equation}
X(t) \equiv \frac{1}{N}\mathbb{E}[\tr(e^{itH(\mathbf{x})})],
\end{equation}
whose Fourier transform directly yields the density of states. Following the previous section, we develop a Dyson series for this representation in powers of $\gamma$:
\begin{equation}
	X(t) = X_0(t)+X_1(t)+X_2(t)+\cdots,
\end{equation}
where
\begin{equation}\label{eq:2ndorderX}
	X_0(t) = \frac{1}{N}\tr(e^K)  \underset{N\rightarrow \infty}{\longrightarrow}e^{2it-\frac{\gamma^2 t^2}{2}}J_0(2t),
\end{equation}
$J_0(t)$ is the Bessel function of the first kind, $K \equiv it T  -\frac{\gamma^2 t^2}{2} \mathbb{I}$, $X_1(t) = 0$,
and $X_2(t)$ is a complicated expression. We can express $X_2(t)$ as a sum of two integrals, $\text{(i)} + \text{(ii)}$, where $\text{(i)} = \frac{\gamma^2t^2}{2} X_0(t)$
and
\begin{multline}
	\text{(ii)} = \gamma^2t^2 \frac{1}{N}\sum_{j=1}^N \int_0^1ds_1 \int_0^{s_1}  ds_2\,\langle j|e^{(1-s_1+s_2)K}|j\rangle \times \\ \langle j|e^{(s_1-s_2)K}|j\rangle.
\end{multline}
To go further we must diagonalise $K$. This is achieved upon introducing the eigenvectors of $K$, $|W_l\rangle \equiv \frac{1}{\sqrt{N}}\sum_{k=1}^{N} e^{\frac{2\pi i}{N}kl}|k\rangle$, which have the corresponding eigenvalues: $\omega_l = i\left(2-2\cos\left(\frac{2\pi}{N}l\right)\right)t - \frac{\gamma^2 t^2}{2}$. Using these observations we have (see the supplementary material), in the limit $N\rightarrow \infty$:  
\begin{equation}
	\text{(ii)} = \gamma^2t  e^{2it-\frac{\gamma^2 t^2}{2}} \frac{\sin(2t)}{4}.
\end{equation}
Similarly, to second order in $\gamma$, we find:
\begin{equation}
	X(t) = e^{2it-\frac{\gamma^2 t^2}{2}}\left[\left(1+\frac{\gamma^2t^2}{2}\right)J_0(2t) - \frac{\gamma^2t}{4}  \sin(2t)\right].
\end{equation}
Taking the Fourier transform of this solution gives us the density of states (this is calculated in the Supplementary material). We see that the diffusion correction gives a simple convolution of the DOS for the tight-binding model with a Gaussian of width $\gamma$. The second-order corrections incorporate the effects of level repulsion. We depict the quantity $X(t)$, calculated to second order, in Fig.~\ref{fig:x} instead of the DOS as it is easier to see the difference between the numerical solution and the approximation in the temporal domain. 

\begin{figure}
	\includegraphics{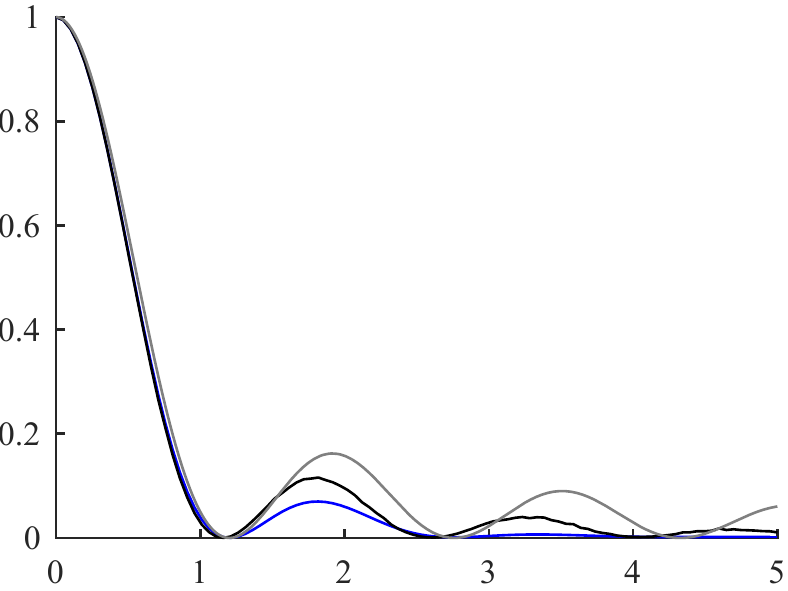}
	\caption{The spectral form factor $\frac{1}{N^2}\mathbb{E}[|\tr(e^{itH})|^2]$ for the Anderson model on 30 sites (the $x$ axis is time in units where $\hbar = 1$). Again, the results for $30$ sites are indistinguishable from the $N=\infty$ limit. Shown black is the result of numerical sampling with $100$ samples. Shown blue is the zeroth order diffusion-corrected term from the stochastic Dyson series. Shown for comparison, in grey, is the zeroth order result from ordinary time-dependent perturbation theory. Note that the zeroth-order stochastic Dyson series result already incorporates the dephasing decay resulting from the disorder average.}\label{fig:sff}
\end{figure}

\paragraph{Application 2: spectral form factors and out-of-time-order correlation functions.---}\hspace{-1em}Here we approximate the $k=1$ spectral form factor Eq.~(\ref{eq:sff}) and the OTOC Eq.~(\ref{eq:otoc}) for the Anderson model. This is expedited by noting that the spectral form factor may be calculated from
$S_{(2)}(t)$, the average propagator for two copies of the Hamiltonian, $H_{(2)} = H_1\otimes \mathbb{I}_{2} - \mathbb{I}_1\otimes H_2$. The OTOC is equal to 
\begin{equation}
	\tr\left(A\otimes B\otimes C\otimes D S_{(4)}(t) \textsc{swap}_{1234} \right),
\end{equation}
where $S_{(4)}(t)$ is the average propagator for four copies of the Hamiltonian, $H_{(4)} = H_1\otimes \mathbb{I}_{234} - \mathbb{I}_1\otimes H_2\otimes \mathbb{I}_{34} + \mathbb{I}_{12}\otimes H_3\otimes \mathbb{I}_{4} - \mathbb{I}_{123}\otimes H_4$, and $\textsc{swap}_{1234}$ is the unitary permutation that cycles the four copies. We can now directly apply the integral representation to both $S_{(2)}(t)$ and $S_{(4)}(t)$ and derive the diffusion approximations:
\begin{equation}
	S_{(2k)}(t) = e^{it (H_0)_{(2k)} - \frac{\gamma^2 t^2}{2} \sum_{j=1}^N (D_j)_{(2k)}^2}.
\end{equation}
Choosing the initial local perturbation to be $A = C = |1\rangle\langle 1|$ and the observation site to be $B = D = |\ell\rangle\langle \ell|$ yields 
\begin{equation}
	\langle 0\ell0\ell|S_{(4)}(t)|\ell0\ell 0\rangle
\end{equation}
for the OTOC. We have depicted the spectral form factor in Fig~\ref{fig:sff}. The OTOC is numerically more intensive and will be the subject of a forthcoming study.

\paragraph{Conclusions and further directions.---}\hspace{-1em}In this Letter we have introduced an exact representation for the disorder-averaged propagator of a quantum system in terms of a stochastic integral over a time-ordered operator expression which involves a Lindblad generator and a temporally random external field, with the path measure given by the Brownian bridge. This expression was then expanded in a stochastic Dyson series to yield a power series in the disorder parameter whose $O(1)$ term explicitly includes diffusive disorder effects. Fluctuations around the diffusive solution arise from the higher-order terms. The representation was exploited to calculate the density of states, spectral form factor, and OTOC for the Anderson model. Much further work remains to be done, including investigating the higher-order corrections, applying tensor-network numerical methods, studying moments, and applications to MBL. This will be the subject of several future papers.

\paragraph{Acknowledgments.---}\hspace{-1em}This work was supported by the DFG through SFB 1227 (DQ-mat), the RTG 1991, and the Max Planck Society through the IMPRS-QST.

\widetext
\section{Supplementary material}

\section{Disorder average of the propagator for an arbitrary impurity model}
In this section we exemplify the general method employed to obtain disorder averages of dynamical processes for disordered systems. The prototype system we consider has a Hamiltonian of the type
\begin{equation}
  H(x) = A+xB,
\end{equation}
where $A$ and $B$ are $D\times D$ matrices and $x\in \mathbb{R}$. The parameter $x$ is chosen randomly according to the Gaussian distribution
\begin{equation}
  g_\gamma(x) \equiv \frac{e^{-\frac{x^2}{2\gamma^2}}}{\sqrt{2\pi}\gamma},
\end{equation}
with variance $\gamma^2$.

We want to be able to calculate disorder averages of all kinds of things. To begin, we'll focus on a simple nontrivial case, namely the disorder-averaged propagator:
\begin{equation}
  S_\gamma(t) \equiv \int g_\gamma(x) e^{it(A+xB)}\, dx.
\end{equation} 
This is, generally, hard to do. One approach to take is to sample $x$ many times and form the empirical average. This works well enough for this example, but scales badly as we reach bigger systems. Also, the large-time limit is noisy.

We are going to need to complete the square numerous times. To this end, we record the following formula:
\begin{equation}
	-\alpha x^2 + \beta x = -\left(\sqrt{\alpha}x-\frac{\beta}{2\sqrt{\alpha}}\right)^2 + \frac{\beta^2}{4{\alpha}}.
\end{equation}

Before we describe our approach we will evaluate $S_{\gamma}(t)$ in the case where $[A,B]=0$, as this is rather instructive; we find
\begin{equation}
	S_\gamma(t) = e^{itA}e^{-\frac{\gamma^2 t^2}{2}B^2}.
\end{equation}

The first step is to break up the evolution into small pieces:
\begin{equation}
  S_\gamma(t) \equiv \int g_\gamma(x) \left(e^{\frac{it}{n}(A+xB)}\right)^n\, dx.
\end{equation}
The next step is to introduce $n$ independent variables $x_1, x_2, \ldots, x_n$ and enforce equality via delta functions:
\begin{equation}
  S_\gamma(t) = \frac{1}{\sqrt{2\pi}\gamma}\int  e^{-\frac{x_1^2}{2n\gamma^2}}\cdots e^{-\frac{x_n^2}{2n\gamma^2}}\delta(x_2-x_1)\cdots \delta(x_n-x_{n-1}) \left(\prod_{j=1}^n e^{\frac{it}{n}(A+x_jB)}\right)\, dx_1\cdots dx_n.
\end{equation}
We eliminate the delta functions by employing the formula
\begin{equation}
	\delta(x) \equiv \frac{1}{2\pi}\int_{-\infty}^\infty e^{ikx}\, dk,
\end{equation}
and find:
\begin{equation}
  S_\gamma(t) = \frac{1}{(2\pi)^{n-1}}\frac{1}{\sqrt{2\pi}\gamma}\int  e^{-\frac{\|\mathbf{x}\|^2}{2n\gamma^2}}e^{ik_1(x_2-x_1)}\cdots e^{ik_{n-1}(x_n-x_{n-1})} \left(\prod_{j=1}^n e^{\frac{it}{n}(A+x_jB)}\right)\, dx_1\cdots dx_ndk_1\cdots dk_{n-1}.
\end{equation}
Introducing two auxiliary variables, $k_0=k_n=0$, we find:
\begin{equation}
  S_\gamma(t) = \frac{1}{(2\pi)^{n-1}}\frac{1}{\sqrt{2\pi}\gamma}\int  e^{-\frac{\|\mathbf{x}\|^2}{2n\gamma^2}}e^{ix_1(k_0-k_1)}\cdots e^{ix_{n}(k_{n-1}-k_{n})} \left(\prod_{j=1}^n e^{\frac{it}{n}(A+x_jB)}\right)\, dx_1\cdots dx_ndk_1\cdots dk_{n-1}.
\end{equation}
Consider the expression
\begin{equation}
	e^{-\frac{{x}_j^2}{2n\gamma^2}}e^{ix_j(k_{j-1}-k_j)}. 
\end{equation}
In the exponent we use:
\begin{equation}
	\alpha = \frac{1}{2n\gamma^2}, \quad \text{and}\quad \beta = i(k_{j-1}-k_j).
\end{equation}
Complete the square to reexpress it as:
\begin{equation}
	e^{-\frac{{x}_j^2}{2n\gamma^2}}e^{ix_j(k_{j-1}-k_j)} = e^{-\left(\frac{1}{\sqrt{2n}\gamma}x_j-i\sqrt{\frac{n}{2}}\gamma(k_{j-1}-k_j)\right)^2}e^{-\frac{n\gamma^2}{2}(k_{j-1}-k_j)^2}.
\end{equation}
Substitute this value into $S_\gamma(t)$:
\begin{equation}
  S_\gamma(t) = \frac{1}{(2\pi)^{n-1}}\frac{1}{\sqrt{2\pi}\gamma}\int   e^{-\frac{n\gamma^2}{2}\sum_{j=1}^{n}(k_{j-1}-k_j)^2} \left(\prod_{j=1}^n \widetilde{F}_j(k_{j-1}-k_j)\right)\, dk_1\cdots dk_{n-1},
\end{equation}
where 
\begin{equation}
	\widetilde{F}_j(k_{j-1}-k_j) = \int_{-\infty}^\infty e^{-\left(\frac{1}{\sqrt{2n}\gamma}x_j-i\sqrt{\frac{n}{2}}\gamma(k_{j-1}-k_j)\right)^2} e^{\frac{it}{n}(A+x_jB)}\, dx_j.
\end{equation}

We turn to the calculation of the $\widetilde{F}$ operators. We expand the values in powers of $1/n$ and retain only the terms to $O(n^{-1})$:
\begin{equation}
	\widetilde{F}_j(k_{j-1}-k_j) = \int_{-\infty}^\infty e^{-\frac{1}{{2n}\gamma^2}\left(x_j-in\gamma^2(k_{j-1}-k_j)\right)^2} e^{\frac{it}{n}(A+x_jB)}\, dx_j.
\end{equation}
Then we introduce
\begin{equation}
	\kappa_j \equiv n(k_{j-1}-k_j),
\end{equation}
(this will later become $-\tfrac{dW}{ds}$), so that
\begin{equation}
	\widetilde{F}_j(k_{j-1}-k_j) = \int_{-\infty}^\infty e^{-\frac{1}{{2n}\gamma^2}\left(x_j-i\gamma^2\kappa_j\right)^2} e^{\frac{it}{n}(A+x_jB)}\, dx_j.
\end{equation}
The next step is to expand the exponential in its powers:
\begin{equation}
	 \widetilde{F}_j(k_{j-1}-k_j) \approx \int_{-\infty}^\infty e^{-\frac{1}{{2n}\gamma^2}\left(x_j-i\gamma^2\kappa_j\right)^2} \left(\mathbb{I}+\frac{it}{n}(A+x_j B) - \frac{t^2}{2n^2}(A+x_jB)^2+\cdots\right)\, dx_j.
\end{equation} 
The $O(1)$ term is
\begin{equation}
	 \int_{-\infty}^\infty e^{-\frac{1}{{2n}\gamma^2}\left(x_j-i\gamma^2\kappa_j\right)^2} \, dx_j =  \int_{-\infty}^\infty e^{-\frac{1}{{2n}\gamma^2} x_j^2} \, dx_j = \sqrt{2\pi n} \gamma.
\end{equation} 

The first-order contribution is calculated using 
\begin{equation}
	 \int_{-\infty}^\infty x_j e^{-\frac{1}{{2n}\gamma^2}\left(x_j-i\gamma^2\kappa_j\right)^2} \, dx_j =  \int_{-\infty}^\infty (x_j+i\gamma^2\kappa_j) e^{-\frac{1}{{2n}\gamma^2} x_j^2} \, dx_j = i\gamma^2\kappa_j\sqrt{2\pi n} \gamma.
\end{equation} 
Thus, we find:
\begin{equation}
	 \frac{it}{n}\int_{-\infty}^\infty e^{-\frac{1}{{2n}\gamma^2}\left(x_j-i\gamma^2\kappa_j\right)^2}(A+x_j B) \, dx_j =  \frac{it}{n}\sqrt{2\pi n}\gamma \left(A+i\gamma^2\kappa_j B\right).
\end{equation} 

The second-order contribution may be calculated using
\begin{equation}
	\begin{split}
		\int_{-\infty}^\infty x_j^2 e^{-\frac{1}{{2n}\gamma^2}\left(x_j-i\gamma^2\kappa_j\right)^2} \, dx_j =  \int_{-\infty}^\infty (x_j+i\gamma^2\kappa_j)^2 e^{-\frac{1}{{2n}\gamma^2} x_j^2} \, dx_j &= \int_{-\infty}^\infty x_j^2 e^{-\frac{1}{{2n}\gamma^2} x_j^2} \, dx_j -\gamma^4\kappa_j^2 \sqrt{2\pi n}\gamma \\
		&= (n\gamma^2 -\gamma^4\kappa_j^2) \sqrt{2\pi n}\gamma.
	\end{split}
\end{equation} 
Then we calculate
\begin{equation}
	 -\frac{t^2}{2n^2}\int_{-\infty}^\infty e^{-\frac{1}{{2n}\gamma^2}\left(x_j-i\gamma^2\kappa_j\right)^2}(A+x_j B)^2 \, dx_j = -\frac{t^2}{2n^2}\int_{-\infty}^\infty e^{-\frac{1}{{2n}\gamma^2}\left(x_j-i\gamma^2\kappa_j\right)^2}(A^2+2x_j \{A,B\} + x_j^2B^2) \, dx_j
\end{equation} 
and find that only the $x_j^2B^2$ term is of the order of $t/n$:
\begin{equation}
	 -\frac{t^2}{2n^2}\int_{-\infty}^\infty e^{-\frac{1}{{2n}\gamma^2}\left(x_j-i\gamma^2\kappa_j\right)^2}(A+x_j B)^2 \, dx_j = -\frac{t^2}{2n^2} B^2(n\gamma^2 -\gamma^4\kappa_j^2) \sqrt{2\pi n}\gamma .
\end{equation} 

Putting this all together, we find:
\begin{equation}
	\widetilde{F}_j(k_{j-1}-k_j) \approx  \sqrt{2\pi n}\gamma\left(\mathbb{I} + \frac{it}{n}(A+i\gamma^2n(k_{j-1}-k_j) B ) - \frac{t^2}{2n^2}(n\gamma^2 - n^2\gamma^4(k_{j-1}-k_j)^2 )B^2\right).
\end{equation}

It is convenient to take out the overall factor $\sqrt{2\pi n}\gamma$ and define
\begin{equation}
	F_j(k_{j-1}-k_j) = \frac{\widetilde{F}_j(k_{j-1}-k_j)}{\sqrt{2\pi n}\gamma}.
\end{equation}

Now that we have the formula for the $F$ and $\widetilde{F}$ operators, we can put together the normalisations and we find:
\begin{equation}
  S_\gamma(t) = \frac{1}{(2\pi)^{n-1}}\frac{1}{\sqrt{2\pi}\gamma} (2\pi n \gamma^2)^{\frac{n}{2}}\int   e^{-\frac{n\gamma^2}{2}\sum_{j=1}^{n}(k_{j-1}-k_j)^2} \left(\prod_{j=1}^n F_j(k_{j-1}-k_j)\right)\, dk_1\cdots dk_{n-1}.
\end{equation}
Let's work out the normalisation required to make the integral over the $k$s a probability measure. To this end we write the exponent in matrix form:
\begin{equation}
	e^{-\frac{n\gamma^2}{2}\sum_{j=1}^{n}(k_{j-1}-k_j)^2} = e^{-\frac{n\gamma^2}{2}\mathbf{k}^T \mathbf{M} \mathbf{k}},
\end{equation}
where $\mathbf{k} = (k_1, \ldots, k_{n-1})$ and the $(n-1)\times (n-1)$ matrix $\mathbf{M}$ is given by
\begin{equation}
	\mathbf{M} = \begin{pmatrix}
		2  & -1 & 0  & 0   &\cdots & 0  \\
		-1 & 2  & -1 & 0   &\cdots & 0  \\
		0  & -1 & 2  & -1  &       & 0    \\
		\vdots   &    &    &     &\ddots & \vdots      \\
		0  &  \cdots  &    &	0   &    -1   & 2  
	\end{pmatrix} =  2\mathbb{I} -P_{n-1}.
\end{equation}
$P_{n-1}$ is the adjacency matrix of the \emph{path graph}. The matrix $\mathbf{M}$ can be diagonalised, which gives the eigenvalues
\begin{equation}
	\lambda_{j} = 2-2\cos\left(\frac{\pi j}{n}\right), \quad j = 1,2, \ldots, n-1.
\end{equation}
The corresponding eigenvector is given by
\begin{equation}
	\mathbf{v}_j = \left(\frac12\sin\left(\frac{\pi j}{n}\right), \frac12\sin\left(\frac{2\pi j}{n}\right), \ldots, \frac12\sin\left(\frac{(n-1)\pi j}{n}\right)\right).
\end{equation}
(These eigenvectors need to be normalised; the normalisation of $\mathbf{v}_j$ is $\sqrt{(n)/8}$.)
We've also gathered, from the Laplace expansion (twice), that $\det(\mathbf{M}) = n$.

We need the gaussian integral formula
\begin{equation}
	\int e^{-\frac12 \mathbf{x}^T\mathbf{A}\mathbf{x}}\, dx_1\cdots dx_n = \sqrt{\frac{(2\pi)^n}{\det \mathbf{A}}}.
\end{equation}
In our case, we have
\begin{equation}
	\int e^{-\frac{n\gamma^2}{2}\mathbf{k}^T \mathbf{M} \mathbf{k}} \, dk_1\cdots dk_{n-1} = \sqrt{\frac{(2\pi)^{n-1}}{n(n\gamma^2)^{n-1}}},
\end{equation}
which means that
\begin{equation}
	d\mu \equiv \sqrt{\frac{n(n\gamma^2)^{n-1}}{(2\pi)^{n-1}}} e^{-\frac{n\gamma^2}{2}\mathbf{k}^T \mathbf{M} \mathbf{k}} \, dk_1\cdots dk_{n-1}  
\end{equation}
is a probability measure.

Putting this together, we find:
\begin{equation}
  S_\gamma(t) = \frac{1}{(2\pi)^{n-1}}\frac{1}{\sqrt{2\pi}\gamma} (2\pi n \gamma^2)^{\frac{n}{2}} \sqrt{\frac{(2\pi)^{n-1}}{n(n\gamma^2)^{n-1}}} \int   \left(\prod_{j=1}^n F_j(k_{j-1}-k_j)\right)\, d\mu = \int   \left(\prod_{j=1}^n F_j(k_{j-1}-k_j)\right)\, d\mu.
\end{equation}

The next step is to identify the measure $d\mu$. Our contention is that it is the (discretised) Brownian bridge measure. Before we do this, we make a basic consistency check: suppose $B=0$; then we have that $F \approx e^{i\frac{t}{n}A}$, so $S_\gamma(t) = e^{itA}$, as it should. The other case we can directly calculate is for $A=0$. Using the exponential form, 
\begin{equation}
  F_j(k_{j-1}-k_j) \approx e^{\frac{it}{n}\left(A+in\gamma^2(k_{j-1}-k_j)B\right) - \frac{t^2}{2n}\gamma^2B^2},
\end{equation}
we find, in the case of $A=0$,
\begin{equation}
  F_j(k_{j-1}-k_j) \approx e^{-t\gamma^2(k_{j-1}-k_j)B - \frac{t^2}{2n}\gamma^2B^2}.
\end{equation}
Substituting this into $S_\gamma(t)$ we find:
\begin{equation}
  S_\gamma(t) = \int\left(\prod_{j=1}^n F_j(k_{j-1}-k_j)\right)\, d\mu = \int \prod_{j=1}^n\left( e^{-t\gamma^2(k_{j-1}-k_j)B - \frac{t^2}{2n}\gamma^2B^2}\right)\, d\mu.
\end{equation}
The sum in the exponential collapses to zero and we are left with
\begin{equation}
  S_\gamma(t) = e^{-\frac{\gamma^2 t^2}{2}B^2},
\end{equation}
as required.

The next stage of our argument is to realise $S_\gamma(t)$ as the expectation value of the operator
\begin{equation}
  \prod_{j=1}^n F_j(k_{j-1}-k_j) \approx \prod_{j=1}^n e^{\frac{it}{n}\left(A+in\gamma^2(k_{j-1}-k_j)B\right) - \frac{t^2}{2n}\gamma^2B^2} \equiv U(\mathbf{k})
\end{equation}
over random paths $\mathbf{k} \equiv (k_0=0, k_1, \ldots, k_{n-1}, k_n=0)$, sampled according to the \emph{path measure} $d\mu$.

We will eventually identify these paths as coming from the \emph{Brownian bridge}. This requires several steps; we begin by first discussing the discretisation of Brownian motion and then move onto the bridge. 

Recall that a Brownian motion or \emph{Wiener process} $W_t$ is characterised by the following four properties:
\begin{enumerate}
	\item $W_0= 0$;
	\item $W_t$ is (almost surely) continuous;
	\item $W_t$ has independent increments; and
	\item $W_t-W_s$ is distributed according to the normal distribution with zero mean and variance $t-s$ for $0\le s\le t$.
\end{enumerate}
This characterisation provides us with a recipe to approximate $W_t$: first discretise the interval $[0,t]$ into $n$ subintervals:
\begin{equation}
	[0,t] = [0, \tfrac{t}{n})\cup [\tfrac{t}{n},\tfrac{2t}{n}) \cup \cdots \cup [\tfrac{(n-1)t}{n},t].
\end{equation}
Define
\begin{equation}
	\Delta w_j \equiv W_{\frac{jt}{n}} - W_{\frac{(j-1)t}{n}}.
\end{equation}
According to the fourth property, we know that $\Delta w_j$ is distributed according the normal distribution with the probability distribution function
\begin{equation}
	p_{t,n}(x) \equiv \sqrt{\frac{n}{2\pi t}} e^{-\frac{n}{t}\frac{x^2}{2}}.
\end{equation}
We also have that
\begin{equation}
	W_t = \sum_{j=1}^n \Delta w_j.
\end{equation}
The probability distribution function for our discretisation is hence
\begin{equation}
	\sqrt{\frac{n^n}{(2\pi t)^n}} e^{-\frac{n}{t}\sum_{j=1}^n\frac{(\Delta w_j)^2}{2}}.
\end{equation}
Let's change variables from $\Delta w_j$ to $w_j \equiv \sum_{k=1}^j \Delta w_k$. The Jacobian for this change of variables has the matrix elements:
\begin{equation}
	[\mathbf{J}]_{jk} \equiv \frac{\partial w_j}{\partial \Delta w_k} = \begin{cases}
		1, & \quad k \le j, \\
		0, & \quad \text{otherwise,}
	\end{cases}
\end{equation}
so that $\mathbf{J}$ is the lower-triangular matrix with $1$s in all the nonzero entries:
\begin{equation}
	\mathbf{J} \equiv \begin{pmatrix}
	1 & 0 & 0 & \cdots & 0 \\
	1 & 1 & 0 & \cdots & 0 \\
	\vdots & \vdots & & \ddots & \vdots \\
	1 & 1 & 1 & \cdots & 1
	\end{pmatrix}.
\end{equation}
The inverse matrix $\mathbf{J}^{-1}$ is given by
\begin{equation}
	\mathbf{D} \equiv \mathbf{J}^{-1} \equiv \begin{pmatrix}
	1 & 0 & 0 & \cdots & 0 \\
	-1 & 1 & 0 & \cdots & 0 \\
	0 & -1 & 1 & \cdots & 0 \\
	\vdots & \vdots & & \ddots & \vdots \\
	0 & 0 & \cdots & -1 & 1
	\end{pmatrix}.
\end{equation}
Thus we can express the probability density function for $w_j$ as 
\begin{equation}
	\sqrt{\frac{n^n}{(2\pi t)^n}} e^{-\frac{n}{t}\frac{\mathbf{w}^T \mathbf{D}^T\mathbf{D} \mathbf{w}}{2}}.
\end{equation}
According to Donsker's theorem, the continuum limit ($N\rightarrow \infty$) of this construction tends (in distribution) to $W_t$. Note that the matrix $\mathbf{D}^T\mathbf{D}$ is a triangular matrix of the form
\begin{equation}
	\mathbf{D}^T\mathbf{D} = \begin{pmatrix}
	2 & -1 & 0 & \cdots & 0 \\
	-1 & 2 & -1 & \cdots & 0 \\
	0 & -1 & 2 & \cdots & 0 \\
	\vdots & \vdots & & \ddots & \vdots \\
	0 & 0 & \cdots & 2 & -1 \\
	0 & 0 & \cdots & -1 & 1
	\end{pmatrix}.
\end{equation}

The path measure for $k_j$ is, however, slightly different in a crucial way. To understand this measure, we introduce the variables
\begin{equation}
	b_j \equiv w_j - \frac{j}{n} w_n, \quad j = 0, 1, \ldots, n.
\end{equation}
These variables are the discretisation of the \emph{Brownian bridge} $B_t$, which is a continuous-time stochastic process with the same conditional probability distribution as the Wiener process, but subject to the condition that at $t=1$ it is pinned to $0$, i.e., $B_1 = 0$. The Brownian bridge is defined by
\begin{equation}
	B_t = W_t - tW_{t=1}.
\end{equation}
Note that the increments of the Brownian bridge are not independent.

These random variables have the property that $b_0 = b_n = 0$. In terms of the variables $\Delta w_j$ we have:
\begin{equation}
	b_j = \sum_{k=1}^{j} \Delta w_k - \frac{j}{n} \sum_{k=1}^{n} \Delta w_k, \quad j = 0, 1, \ldots, n.
\end{equation}
The Jacobian relating $b$ with $\Delta w_j$ has matrix elements
\begin{equation}
	[\Gamma^{-1}]_{jk} \equiv \frac{\partial b_j}{\partial \Delta w_k} = \begin{cases}
		1 - \frac{j}{n}, & \quad k \le j, \\
		-\frac{j}{n}, & \quad \text{otherwise.}
		\end{cases}
\end{equation}
This matrix has the form
\begin{equation}
	\Gamma^{-1} = \begin{pmatrix}
	1-\frac{1}{n} & -\frac1n & -\frac1n & \cdots & -\frac1n \\[0.2em]
	1-\frac{2}{n} & 1-\frac{2}{n} & -\frac2n & \cdots & -\frac2n \\[0.2em]
	1-\frac{3}{n} & 1-\frac{3}{n} & 1-\frac3n & \cdots & -\frac3n \\[0.2em]
	\vdots & \vdots & & \ddots & \vdots \\[0.2em]
	\frac{1}{n} & \frac{1}{n} & \frac{1}{n} & \cdots & -1+\frac{1}n \\[0.2em]
	0 & 0 & 0 & \cdots & 0
	\end{pmatrix}.
\end{equation}
The matrix $\Gamma$ (for which $\Gamma^{-1}$ is the partial \emph{left inverse}) is then given by
\begin{equation}
	\Gamma \equiv \begin{pmatrix}
	1 & 0 & 0 & 0 &\cdots & 0 \\
	-1 & 1 & 0 & 0 &\cdots & 0 \\
	0 & -1 & 1 & 0 &\cdots & 0 \\
	\vdots & \vdots &\vdots & & \ddots & \vdots \\
	0 & \cdots & 0 & -1 & 1 & 0 \\
	0 & \cdots & 0 & 0 & -1 & 0
	\end{pmatrix}.
\end{equation}
Since the $(n-1)\times (n-1)$ matrix $\mathbf{M}$ is determined by the $(n-1)\times (n-1)$ submatrix of $\Gamma^T\Gamma$, we have that its inverse $\mathbf{M}^{-1}$ is given by the corresponding $(n-1)\times (n-1)$ submatrix of $\Gamma^{-1}(\Gamma^{-1})^T$. Calculating the matrix elements, we have:
\begin{equation}
	[\mathbf{M}^{-1}]_{j,k} = \min\left\{j\left(1-\tfrac{k}{n}\right), k\left(1-\tfrac{j}{n}\right)\right\}.
\end{equation}
Now that we have a formula for the inverse of $\mathbf{M}$, we are able to calculate moments via the generating function:
\begin{equation}
	 	\sqrt{\frac{n(n\gamma^2)^{n-1}}{(2\pi)^{n-1}}} \int e^{-\frac{n\gamma^2}{2}\mathbf{k}^T \mathbf{M} \mathbf{k}}e^{\boldsymbol{\ell}^T\mathbf{k}} \, dk_1\cdots dk_{n-1} = e^{\frac{1}{2n\gamma^2}\boldsymbol{\ell}^T\mathbf{M}^{-1}\boldsymbol{\ell}}.
\end{equation} 
Thus we obtain, e.g., for $j\le j'$:
\begin{equation}
	\langle k_jk_{j'} \rangle = \frac{\partial}{\partial \ell_j}\frac{\partial}{\partial \ell_k} e^{\frac{1}{2n\gamma^2}\boldsymbol{\ell}^T\mathbf{M}^{-1}\boldsymbol{\ell}} \bigg|_{\boldsymbol{\ell}=\mathbf{0}} = \frac{1}{n\gamma^2}[\mathbf{M}^{-1}]_{j,j'} = \frac{1}{\gamma^2}\frac{j}{n}\left(1-\frac{j'}{n}\right).
\end{equation}	

\subsection{The continuum limit}
Taking the continuum limit will yield a coupled set of stochastic differential equations (SDE). Our starting point is the expression
\begin{equation}
  S_\gamma(t) = \sqrt{\frac{n(n\gamma^2)^{n-1}}{(2\pi)^{n-1}}}  \int   e^{-\frac{n\gamma^2}{2}\mathbf{k}^T \mathbf{M} \mathbf{k}}\left(\prod_{j=1}^n F_j(k_{j-1}-k_j)\right)\, dk_1\cdots d{k_{n-1}}.
\end{equation}
We first scale out $\gamma$ by defining $l_j = \gamma k_j$; we obtain:
\begin{equation}
  S_\gamma(t) = \sqrt{\frac{n^{n}}{(2\pi)^{n-1}}}  \int   e^{-\frac{n}{2}\mathbf{l}^T \mathbf{M} \mathbf{l}}\left(\prod_{j=1}^n F_j\left(\frac{1}{\gamma}(l_{j-1}-l_j)\right)\right)\, dl_1\cdots d{l_{n-1}}.
\end{equation}
Thus, by substituting for $F_j$ it follows:
\begin{equation}
	S_\gamma(t) = \sqrt{\frac{n^{n}}{(2\pi)^{n-1}}}  \int   e^{-\frac{n}{2}\mathbf{l}^T \mathbf{M} \mathbf{l}}\prod_{j=1}^n \left(\mathbb{I} + \frac{it}{n}A+t\gamma (\Delta l_j) B  - \frac{t^2}{2n^2}(n\gamma^2 - n^2\gamma^2(\Delta l_j)^2 )B^2\right)\, dl_1\cdots d{l_{n-1}},
\end{equation}
where
\begin{equation}
	\Delta l_j = l_{j}-l_{j-1}.
\end{equation}
As we explained earlier, $l_j$ may be identified with a discretisation of the standard Brownian bridge. 

We define 
\begin{equation}
	X_k \equiv \prod_{j=1}^k \left(\mathbb{I} + \frac{it}{n}A+t\gamma (\Delta l_j) B  - \frac{t^2}{2n^2}(n\gamma^2 - n^2\gamma^2(\Delta l_j)^2 )B^2\right).
\end{equation}
Using this expression we have:
\begin{equation}
	X_{k+1} = \left(\mathbb{I} + \frac{it}{n}A+t\gamma (\Delta l_k) B  - \frac{t^2}{2n^2}(n\gamma^2 - n^2\gamma^2(\Delta l_k)^2 )B^2\right)X_k.
\end{equation}
The difference between $X_{k+1}-X_k$ is thus
\begin{equation}
	\Delta X_{k} \equiv X_{k+1}-X_k = \left(\frac{it}{n}A+t\gamma (\Delta l_k) B  - \frac{t^2}{2n^2}(n\gamma^2 - n^2\gamma^2(\Delta l_k)^2 )B^2\right)X_k.
\end{equation}
We define
\begin{equation}
	dz_{s=k/n} = \Delta l_k \frac{t}{n}
\end{equation}
and
\begin{equation}
	X_{s = k/n} \equiv \Delta X_{k},
\end{equation}
and write the increment $\Delta X_{k}$ as $dX_s$ in the limit of $N\rightarrow \infty$. By putting this together we obtain the system of stochastic differential equations:

\begin{equation}\label{eq:ABzsde}
\boxed{		\begin{split}
	dX_s &= it A X_s \,ds + t\gamma B X_s \,dz_s ,\\
		dz_s &= -\frac{z_s}{1-s}\, ds + dW_s.
	\end{split}}
\end{equation}
This is a consequence of the fact that, in the distribution, 
\begin{equation}
	(\Delta l_k)^2 = \frac{1}{n}-\frac{1}{n^2},
\end{equation}
and the second term is negligible in the limit.

\section{Calculations for the Anderson model}
Here we detail the calculations for the Anderson model. We focus on the quantity
\begin{equation}
X(t) \equiv \frac{1}{n}\mathbb{E}[\tr(e^{itH(\mathbf{x})})].
\end{equation}
By employing the stochastic integral representation we have :
\begin{equation}
	X(t) = \frac{1}{n}\int \tr\left(\mathcal{T}e^{\int_0^1 it T  -\frac{\gamma^2 t^2}{2} \mathbb{I}\, ds + \gamma t\sum_{j=1}^n\int_0^1  D_j \,dz_j}\right) \,  d\mu(z) .
\end{equation}
By following the previous section, we develop a Dyson series for this representation in powers of $\gamma$:
\begin{equation}
	X(t) = X_0(t)+X_1(t)+X_2(t)+\cdots,
\end{equation}
where
\begin{equation}
	X_0(t) = \frac{1}{n}\tr(e^K),
\end{equation}
with
\begin{equation}
	K \equiv it T  -\frac{\gamma^2 t^2}{2} \mathbb{I},
\end{equation}
\begin{equation}
	X_1(t) = 0,
\end{equation}
and $X_2(t)$ is a complicated expression. We can express $X_2(t) = \text{(i)} + \text{(ii)}$ in terms of two integrals:
\begin{equation}
	X_2(t) = \frac{\gamma^2t^2}{2} \frac{1}{n}\sum_{j=1}^n\int_0^1 \tr\left(e^{(1-s)K}D_je^{s K}\right)\,ds -\gamma^2t^2 \frac{1}{n}\sum_{j=1}^n \int_0^1ds_1 \int_0^{s_1}ds_2\,  \tr\left(e^{(1-s_1)K}D_je^{(s_1-s_2)K} D_je^{s_2 K}\right).
\end{equation}
The first is simple:
\begin{equation}
	\frac{\gamma^2t^2}{2} \frac{1}{n}\sum_{j=1}^n\int_0^1 \tr\left(e^{(1-s)K}D_je^{s K}\right)\,ds = \frac{\gamma^2t^2}{2} X_0(t).
\end{equation}
We can slightly simplify the second integral:
\begin{equation}
	\text{(ii)} = \gamma^2t^2 \frac{1}{n}\sum_{j=1}^n \int_0^1ds_1 \int_0^{s_1}  ds_2\,\langle j|e^{(1-s_1+s_2)K}|j\rangle\langle j|e^{(s_1-s_2)K}|j\rangle.
\end{equation}
To go further, we must diagonalise $K$. This is achieved upon introducing the eigenvectors of $K$,
\begin{equation}
	|W_l\rangle \equiv \frac{1}{\sqrt{n}}\sum_{k=1}^{n} e^{\frac{2\pi i}{n}kl}|k\rangle,
\end{equation}
which have the corresponding eigenvalues:
\begin{equation}
	\omega_l = i\left(2-2\cos\left(\frac{2\pi}{n}l\right)\right)t - \frac{\gamma^2 t^2}{2}.
\end{equation}
Note that 
\begin{equation}
	\langle j|W_l\rangle = \frac{1}{\sqrt{n}} e^{\frac{2\pi i}{n}kl},
\end{equation}
so that
\begin{equation}
	\langle j|W_l\rangle\langle W_l|j\rangle = \frac{1}{n}.
\end{equation}
Using these observations we have:
\begin{equation}
	\text{(ii)} = \gamma^2t^2 \frac{1}{n^2}\sum_{k_1,k_2=1}^n \int_0^1ds_1 \int_0^{s_1}  ds_2\, e^{(1-s_1+s_2)\omega_{k_1}}e^{(s_1-s_2)\omega_{k_2}}.
\end{equation}
Expanding the exponents:
\begin{equation}
	\text{(ii)} = \gamma^2t^2 e^{2it-\frac{\gamma^2 t^2}{2}} \frac{1}{n^2}\sum_{k_1,k_2=1}^n \int_0^1ds_1 \int_0^{s_1}  ds_2\, e^{2it(1-s_1+s_2)\cos\left(\frac{2\pi}{n}k_1\right)}e^{2it(s_1-s_2)\cos\left(\frac{2\pi}{n}k_2\right)}.
\end{equation}
We now take the limit of $N\rightarrow \infty$ and define
\begin{equation}
	z_1 \equiv \frac{2\pi}{n} k_1, \quad \text{and}\quad z_2 \equiv \frac{2\pi}{n} k_2,
\end{equation}
with
\begin{equation}
	dz_2 = dz_1 \approx \frac{2\pi}{n}\text{;}\quad \frac{1}{n}\sum_{k_1=1}^n \approx \frac{1}{2\pi}\int_0^{2\pi} dz_1.
\end{equation}
In this way we obtain
\begin{equation}
	\text{(ii)} = \gamma^2t^2 e^{2it-\frac{\gamma^2 t^2}{2}} \int_0^1ds_1 \int_0^{s_1}  ds_2\, \left\{\frac{1}{2\pi} \int_0^{2\pi} e^{2it(1-s_1+s_2)\cos\left(z_1\right)}\, dz_1 \times \frac{1}{2\pi} \int_0^{2\pi}e^{2it(s_1-s_2)\cos\left(z_2\right)}\,dz_2\right\}.
\end{equation}
We recognise these integrals as representations for the Bessel function of the first kind $J_0(x)$:
\begin{equation}
	\text{(ii)} = \gamma^2t^2 e^{2it-\frac{\gamma^2 t^2}{2}} \int_0^1ds_1 \int_0^{s_1}  ds_2\, J_0(2t(1-s_1+s_2))J_0(2t(s_1-s_2)).
\end{equation}
Using the series representation,
\begin{equation}
	J_0(x) = \sum_{l=0} \frac{(-1)^l}{2^{2l}(l!)^2}x^{2l},
\end{equation}
and integrating explicitly, one can show that the double integral evaluates to:
\begin{equation}
	\int_0^1ds_1 \int_0^{s_1}  ds_2\, J_0(2t(1-s_1+s_2))J_0(2t(s_1-s_2)) = \frac{\sin(2t)}{4t},
\end{equation}
so we have
\begin{equation}
	\text{(ii)} = \gamma^2t  e^{2it-\frac{\gamma^2 t^2}{2}} \frac{\sin(2t)}{4}.
\end{equation}
Similarly, for $X_0(t)$ we find:
\begin{equation}
	X_0(t) = \frac{1}{n}e^{2it-\frac{\gamma^2 t^2}{2}}\sum_{k=1}^n e^{2it\cos\left(\frac{2\pi}{n}k\right)} = e^{2it-\frac{\gamma^2 t^2}{2}} J_0(2t).
\end{equation}
Thus we have, to second order in $\gamma$:
\begin{equation}
	X(t) = \left(1+\frac{\gamma^2t^2}{2}\right)e^{2it-\frac{\gamma^2 t^2}{2}}J_0(2t) - \frac{\gamma^2t}{4}  e^{2it-\frac{\gamma^2 t^2}{2}} \sin(2t).
\end{equation}

Taking the Fourier transform of $X(t)$ gives us the density of states:
\begin{equation}
	\widehat{X}(k) = \left(1+ \left(\frac{i}{2\pi}\right)^2\frac{\gamma^2}{2} \frac{d^2}{dk^2}\right)\sqrt{2\pi}{\gamma^2}e^{-\frac{\pi^2\gamma^2k^2}{2}}\star\frac{\text{rect}(\tfrac{k}{2}-\tfrac{1}{2\pi})}{\sqrt{1-\pi^2 (k-\tfrac{1}{\pi})^2}} - \left(\frac{i}{2\pi}\frac{d}{dk}\right)\frac{\gamma^2}{8i} \sqrt{2\pi}{\gamma^2}e^{-\frac{\pi^2\gamma^2k^2}{2}}\star (\delta(k-\tfrac{2}{\pi}) - \delta(k)).
\end{equation}


\begin{thebibliography}{29}%
\makeatletter
\providecommand \@ifxundefined [1]{%
 \@ifx{#1\undefined}
}%
\providecommand \@ifnum [1]{%
 \ifnum #1\expandafter \@firstoftwo
 \else \expandafter \@secondoftwo
 \fi
}%
\providecommand \@ifx [1]{%
 \ifx #1\expandafter \@firstoftwo
 \else \expandafter \@secondoftwo
 \fi
}%
\providecommand \natexlab [1]{#1}%
\providecommand \enquote  [1]{``#1''}%
\providecommand \bibnamefont  [1]{#1}%
\providecommand \bibfnamefont [1]{#1}%
\providecommand \citenamefont [1]{#1}%
\providecommand \href@noop [0]{\@secondoftwo}%
\providecommand \href [0]{\begingroup \@sanitize@url \@href}%
\providecommand \@href[1]{\@@startlink{#1}\@@href}%
\providecommand \@@href[1]{\endgroup#1\@@endlink}%
\providecommand \@sanitize@url [0]{\catcode `\\12\catcode `\$12\catcode
  `\&12\catcode `\#12\catcode `\^12\catcode `\_12\catcode `\%12\relax}%
\providecommand \@@startlink[1]{}%
\providecommand \@@endlink[0]{}%
\providecommand \url  [0]{\begingroup\@sanitize@url \@url }%
\providecommand \@url [1]{\endgroup\@href {#1}{\urlprefix }}%
\providecommand \urlprefix  [0]{URL }%
\providecommand \Eprint [0]{\href }%
\providecommand \doibase [0]{http://dx.doi.org/}%
\providecommand \selectlanguage [0]{\@gobble}%
\providecommand \bibinfo  [0]{\@secondoftwo}%
\providecommand \bibfield  [0]{\@secondoftwo}%
\providecommand \translation [1]{[#1]}%
\providecommand \BibitemOpen [0]{}%
\providecommand \bibitemStop [0]{}%
\providecommand \bibitemNoStop [0]{.\EOS\space}%
\providecommand \EOS [0]{\spacefactor3000\relax}%
\providecommand \BibitemShut  [1]{\csname bibitem#1\endcsname}%
\let\auto@bib@innerbib\@empty
\bibitem [{\citenamefont {Anderson}(1958)}]{anderson_absence_1958}%
  \BibitemOpen
  \bibfield  {author} {\bibinfo {author} {\bibfnamefont {P.~W.}\ \bibnamefont
  {Anderson}},\ }\href@noop {} {\bibfield  {journal} {\bibinfo  {journal}
  {Phys. Rev.}\ }\textbf {\bibinfo {volume} {109}},\ \bibinfo {pages} {1492}
  (\bibinfo {year} {1958})}\BibitemShut {NoStop}%
\bibitem [{\citenamefont {Lagendijk}\ \emph {et~al.}(2009)\citenamefont
  {Lagendijk}, \citenamefont {van Tiggelen},\ and\ \citenamefont
  {Wiersma}}]{lagendijkFiftyYearsAnderson2009}%
  \BibitemOpen
  \bibfield  {author} {\bibinfo {author} {\bibfnamefont {A.}~\bibnamefont
  {Lagendijk}}, \bibinfo {author} {\bibfnamefont {B.}~\bibnamefont {van
  Tiggelen}}, \ and\ \bibinfo {author} {\bibfnamefont {D.~S.}\ \bibnamefont
  {Wiersma}},\ }\href {\doibase 10.1063/1.3206091} {\bibfield  {journal}
  {\bibinfo  {journal} {Phys. Today}\ }\textbf {\bibinfo {volume} {62}},\
  \bibinfo {pages} {24} (\bibinfo {year} {2009})}\BibitemShut {NoStop}%
\bibitem [{\citenamefont {Schenk}\ \emph {et~al.}(2008)\citenamefont {Schenk},
  \citenamefont {Bollh\"ofer},\ and\ \citenamefont
  {R\"omer}}]{schenkLargeScaleDiagonalizationTechniques2008}%
  \BibitemOpen
  \bibfield  {author} {\bibinfo {author} {\bibfnamefont {O.}~\bibnamefont
  {Schenk}}, \bibinfo {author} {\bibfnamefont {M.}~\bibnamefont {Bollh\"ofer}},
  \ and\ \bibinfo {author} {\bibfnamefont {R.}~\bibnamefont {R\"omer}},\ }\href
  {\doibase 10.1137/070707002} {\bibfield  {journal} {\bibinfo  {journal} {SIAM
  Rev.}\ }\textbf {\bibinfo {volume} {50}},\ \bibinfo {pages} {91} (\bibinfo
  {year} {2008})}\BibitemShut {NoStop}%
\bibitem [{\citenamefont {Abrahams}\ \emph {et~al.}(1979)\citenamefont
  {Abrahams}, \citenamefont {Anderson}, \citenamefont {Licciardello},\ and\
  \citenamefont {Ramakrishnan}}]{abrahams_scaling_1979}%
  \BibitemOpen
  \bibfield  {author} {\bibinfo {author} {\bibfnamefont {E.}~\bibnamefont
  {Abrahams}}, \bibinfo {author} {\bibfnamefont {P.~W.}\ \bibnamefont
  {Anderson}}, \bibinfo {author} {\bibfnamefont {D.~C.}\ \bibnamefont
  {Licciardello}}, \ and\ \bibinfo {author} {\bibfnamefont {T.~V.}\
  \bibnamefont {Ramakrishnan}},\ }\href@noop {} {\bibfield  {journal} {\bibinfo
   {journal} {Phys. Rev. Lett.}\ }\textbf {\bibinfo {volume} {42}},\ \bibinfo
  {pages} {673} (\bibinfo {year} {1979})}\BibitemShut {NoStop}%
\bibitem [{\citenamefont {Wegner}(1979)}]{wegnerMobilityEdgeProblem1979}%
  \BibitemOpen
  \bibfield  {author} {\bibinfo {author} {\bibfnamefont {F.}~\bibnamefont
  {Wegner}},\ }\href@noop {} {\bibfield  {journal} {\bibinfo  {journal} {Z.
  Phys. B Con. Mat.}\ }\textbf {\bibinfo {volume} {35}},\ \bibinfo {pages}
  {207} (\bibinfo {year} {1979})}\BibitemShut {NoStop}%
\bibitem [{\citenamefont {Altshuler}\ \emph {et~al.}(2003)\citenamefont
  {Altshuler}, \citenamefont {Tognetti},\ and\ \citenamefont
  {Tagliacozzo}}]{altshulerQuantumPhenomenaMesoscopic}%
  \BibitemOpen
  \bibinfo {editor} {\bibfnamefont {B.}~\bibnamefont {Altshuler}}, \bibinfo
  {editor} {\bibfnamefont {V.}~\bibnamefont {Tognetti}}, \ and\ \bibinfo
  {editor} {\bibfnamefont {A.}~\bibnamefont {Tagliacozzo}},\ eds.,\ \href@noop
  {} {\emph {\bibinfo {title} {Quantum {{Phenomena}} in {{Mesoscopic
  Systems}}}}},\ \bibinfo {series} {Proceedings of the International School of
  Physics ``Enrico Fermi''}, Vol.\ \bibinfo {volume} {151}\ (\bibinfo
  {publisher} {IOS Press},\ \bibinfo {address} {The Netherlands},\ \bibinfo
  {year} {2003})\BibitemShut {NoStop}%
\bibitem [{\citenamefont {Lerner}\ \emph {et~al.}(2002)\citenamefont {Lerner},
  \citenamefont {Althsuler}, \citenamefont {Fal'ko},\ and\ \citenamefont
  {Giamarchi}}]{lernerStronglyCorrelatedFermions2002}%
  \BibitemOpen
  \bibinfo {editor} {\bibfnamefont {I.~V.}\ \bibnamefont {Lerner}}, \bibinfo
  {editor} {\bibfnamefont {B.~L.}\ \bibnamefont {Althsuler}}, \bibinfo {editor}
  {\bibfnamefont {V.~I.}\ \bibnamefont {Fal'ko}}, \ and\ \bibinfo {editor}
  {\bibfnamefont {T.}~\bibnamefont {Giamarchi}},\ eds.,\ \href@noop {} {\emph
  {\bibinfo {title} {Strongly {{Correlated Fermions}} and {{Bosons}} in
  {{Low}}-{{Dimensional Disordered Systems}}}}},\ Nato Science Series II:\
  (\bibinfo  {publisher} {Springer},\ \bibinfo {address} {The Netherlands},\
  \bibinfo {year} {2002})\BibitemShut {NoStop}%
\bibitem [{\citenamefont {Efetov}(1997)}]{efetov_supersymmetry_1997}%
  \BibitemOpen
  \bibfield  {author} {\bibinfo {author} {\bibfnamefont {K.}~\bibnamefont
  {Efetov}},\ }\href@noop {} {\emph {\bibinfo {title} {Supersymmetry in
  disorder and chaos}}}\ (\bibinfo  {publisher} {Cambridge University Press},\
  \bibinfo {address} {Cambridge},\ \bibinfo {year} {1997})\BibitemShut
  {NoStop}%
\bibitem [{\citenamefont
  {Zirnbauer}(1996)}]{zirnbauerSupersymmetrySystemsUnitary1996}%
  \BibitemOpen
  \bibfield  {author} {\bibinfo {author} {\bibfnamefont {M.~R.}\ \bibnamefont
  {Zirnbauer}},\ }\href@noop {} {\bibfield  {journal} {\bibinfo  {journal} {J.
  Phys. A}\ }\textbf {\bibinfo {volume} {29}},\ \bibinfo {pages} {7113}
  (\bibinfo {year} {1996})}\BibitemShut {NoStop}%
\bibitem [{\citenamefont {Kamenev}\ and\ \citenamefont
  {Levchenko}(2009)}]{kamenev_keldysh_2009}%
  \BibitemOpen
  \bibfield  {author} {\bibinfo {author} {\bibfnamefont {A.}~\bibnamefont
  {Kamenev}}\ and\ \bibinfo {author} {\bibfnamefont {A.}~\bibnamefont
  {Levchenko}},\ }\href@noop {} {\bibfield  {journal} {\bibinfo  {journal}
  {Adv. Phys.}\ }\textbf {\bibinfo {volume} {58}},\ \bibinfo {pages} {197}
  (\bibinfo {year} {2009})}\BibitemShut {NoStop}%
\bibitem [{\citenamefont {Kamenev}\ and\ \citenamefont
  {Andreev}(1999)}]{kamenev_electron-electron_1999}%
  \BibitemOpen
  \bibfield  {author} {\bibinfo {author} {\bibfnamefont {A.}~\bibnamefont
  {Kamenev}}\ and\ \bibinfo {author} {\bibfnamefont {A.}~\bibnamefont
  {Andreev}},\ }\href@noop {} {\bibfield  {journal} {\bibinfo  {journal} {Phys.
  Rev. B}\ }\textbf {\bibinfo {volume} {60}},\ \bibinfo {pages} {2218}
  (\bibinfo {year} {1999})}\BibitemShut {NoStop}%
\bibitem [{\citenamefont
  {Kamenev}(2011)}]{kamenevFieldTheoryNonEquilibrium2011}%
  \BibitemOpen
  \bibfield  {author} {\bibinfo {author} {\bibfnamefont {A.}~\bibnamefont
  {Kamenev}},\ }\href@noop {} {\emph {\bibinfo {title} {Field {{Theory}} of
  {{Non}}-{{Equilibrium Systems}}}}},\ \bibinfo {edition} {1st}\ ed.\ (\bibinfo
   {publisher} {{Cambridge University Press}},\ \bibinfo {address} {New York},\
  \bibinfo {year} {2011})\BibitemShut {NoStop}%
\bibitem [{\citenamefont {Basko}\ \emph {et~al.}(2006)\citenamefont {Basko},
  \citenamefont {Aleiner},\ and\ \citenamefont
  {Altshuler}}]{baskoMetalInsulatorTransition2006}%
  \BibitemOpen
  \bibfield  {author} {\bibinfo {author} {\bibfnamefont {D.~M.}\ \bibnamefont
  {Basko}}, \bibinfo {author} {\bibfnamefont {I.~L.}\ \bibnamefont {Aleiner}},
  \ and\ \bibinfo {author} {\bibfnamefont {B.~L.}\ \bibnamefont {Altshuler}},\
  }\href@noop {} {\bibfield  {journal} {\bibinfo  {journal} {Ann. Phys.}\
  }\textbf {\bibinfo {volume} {321}},\ \bibinfo {pages} {1126} (\bibinfo {year}
  {2006})}\BibitemShut {NoStop}%
\bibitem [{\citenamefont {Alet}\ and\ \citenamefont
  {Laflorencie}(2018)}]{alet_many-body_2018}%
  \BibitemOpen
  \bibfield  {author} {\bibinfo {author} {\bibfnamefont {F.}~\bibnamefont
  {Alet}}\ and\ \bibinfo {author} {\bibfnamefont {N.}~\bibnamefont
  {Laflorencie}},\ }\href@noop {} {\  (\bibinfo {year} {2018})},\ \Eprint
  {http://arxiv.org/abs/arXiv:1711.03145} {arXiv:1711.03145} \BibitemShut
  {NoStop}%
\bibitem [{\citenamefont {Nandkishore}\ and\ \citenamefont
  {Huse}(2015)}]{nandkishoreManyBodyLocalizationThermalization2015}%
  \BibitemOpen
  \bibfield  {author} {\bibinfo {author} {\bibfnamefont {R.}~\bibnamefont
  {Nandkishore}}\ and\ \bibinfo {author} {\bibfnamefont {D.~A.}\ \bibnamefont
  {Huse}},\ }\href@noop {} {\bibfield  {journal} {\bibinfo  {journal} {Ann.
  Rev. Conden. Ma. P.}\ }\textbf {\bibinfo {volume} {6}},\ \bibinfo {pages}
  {15} (\bibinfo {year} {2015})}\BibitemShut {NoStop}%
\bibitem [{\citenamefont {Altman}\ and\ \citenamefont
  {Vosk}(2015)}]{altmanUniversalDynamicsRenormalization2015}%
  \BibitemOpen
  \bibfield  {author} {\bibinfo {author} {\bibfnamefont {E.}~\bibnamefont
  {Altman}}\ and\ \bibinfo {author} {\bibfnamefont {R.}~\bibnamefont {Vosk}},\
  }\href@noop {} {\bibfield  {journal} {\bibinfo  {journal} {Ann. Rev. Conden.
  Ma. P.}\ }\textbf {\bibinfo {volume} {6}},\ \bibinfo {pages} {383} (\bibinfo
  {year} {2015})}\BibitemShut {NoStop}%
\bibitem [{\citenamefont {Abanin}\ and\ \citenamefont
  {Papi\'c}(2017)}]{abaninRecentProgressManybody2017}%
  \BibitemOpen
  \bibfield  {author} {\bibinfo {author} {\bibfnamefont {D.~A.}\ \bibnamefont
  {Abanin}}\ and\ \bibinfo {author} {\bibfnamefont {Z.}~\bibnamefont
  {Papi\'c}},\ }\href@noop {} {\bibfield  {journal} {\bibinfo  {journal} {Ann.
  Phys.}\ }\textbf {\bibinfo {volume} {529}},\ \bibinfo {pages} {1700169}
  (\bibinfo {year} {2017})}\BibitemShut {NoStop}%
\bibitem [{\citenamefont {Liao}\ \emph {et~al.}(2017)\citenamefont {Liao},
  \citenamefont {Levchenko},\ and\ \citenamefont
  {Foster}}]{liaoResponseTheoryErgodic2017}%
  \BibitemOpen
  \bibfield  {author} {\bibinfo {author} {\bibfnamefont {Y.}~\bibnamefont
  {Liao}}, \bibinfo {author} {\bibfnamefont {A.}~\bibnamefont {Levchenko}}, \
  and\ \bibinfo {author} {\bibfnamefont {M.~S.}\ \bibnamefont {Foster}},\
  }\href@noop {} {\bibfield  {journal} {\bibinfo  {journal} {Ann. Phys.}\
  }\textbf {\bibinfo {volume} {386}},\ \bibinfo {pages} {97} (\bibinfo {year}
  {2017})},\ \Eprint {http://arxiv.org/abs/arXiv:1706.07066} {arXiv:1706.07066}
  \BibitemShut {NoStop}%
\bibitem [{\citenamefont {Kitaev}()}]{breakthroughAlexeiKitaev2015}%
  \BibitemOpen
  \bibfield  {author} {\bibinfo {author} {\bibfnamefont {A.}~\bibnamefont
  {Kitaev}},\ }\href@noop {} {\enquote {\bibinfo {title} {Alexei {{Kitaev}}:
  2015 {{Breakthrough Prize Fundamental Physics Symposium}}},}\ }\bibinfo
  {note} {Https://www.youtube.com/watch?v=OQ9qN8j7EZI}\BibitemShut {NoStop}%
\bibitem [{\citenamefont {Swingle}\ and\ \citenamefont
  {Chowdhury}(2017)}]{swingleSlowScramblingDisordered2017}%
  \BibitemOpen
  \bibfield  {author} {\bibinfo {author} {\bibfnamefont {B.}~\bibnamefont
  {Swingle}}\ and\ \bibinfo {author} {\bibfnamefont {D.}~\bibnamefont
  {Chowdhury}},\ }\href@noop {} {\bibfield  {journal} {\bibinfo  {journal}
  {Phys. Rev. B}\ }\textbf {\bibinfo {volume} {95}} (\bibinfo {year} {2017})},\
  \Eprint {http://arxiv.org/abs/arXiv:1608.03280} {arXiv:1608.03280}
  \BibitemShut {NoStop}%
\bibitem [{\citenamefont {Larkin}\ and\ \citenamefont
  {Ovchinnikov}(1969)}]{larkinQuasiclassicalMethodTheory1969}%
  \BibitemOpen
  \bibfield  {author} {\bibinfo {author} {\bibfnamefont {A.~I.}\ \bibnamefont
  {Larkin}}\ and\ \bibinfo {author} {\bibfnamefont {Y.~N.}\ \bibnamefont
  {Ovchinnikov}},\ }\href@noop {} {\bibfield  {journal} {\bibinfo  {journal}
  {Sov. Phys. JETP}\ }\textbf {\bibinfo {volume} {28}},\ \bibinfo {pages}
  {1200} (\bibinfo {year} {1969})}\BibitemShut {NoStop}%
\bibitem [{\citenamefont {Kos}\ \emph {et~al.}(2018)\citenamefont {Kos},
  \citenamefont {Ljubotina},\ and\ \citenamefont
  {Prosen}}]{kosManyBodyQuantumChaos2018}%
  \BibitemOpen
  \bibfield  {author} {\bibinfo {author} {\bibfnamefont {P.}~\bibnamefont
  {Kos}}, \bibinfo {author} {\bibfnamefont {M.}~\bibnamefont {Ljubotina}}, \
  and\ \bibinfo {author} {\bibfnamefont {T.}~\bibnamefont {Prosen}},\
  }\href@noop {} {\bibfield  {journal} {\bibinfo  {journal} {Phys. Rev. X}\
  }\textbf {\bibinfo {volume} {8}},\ \bibinfo {pages} {021062} (\bibinfo {year}
  {2018})}\BibitemShut {NoStop}
\bibitem [{\citenamefont {Bertini}\ \emph {et~al.}(2018)\citenamefont
  {Bertini}, \citenamefont {Kos},\ and\ \citenamefont
  {Prosen}}]{bertiniExactSpectralForm2018}%
  \BibitemOpen
  \bibfield  {author} {\bibinfo {author} {\bibfnamefont {B.}~\bibnamefont
  {Bertini}}, \bibinfo {author} {\bibfnamefont {P.}~\bibnamefont {Kos}}, \ and\
  \bibinfo {author} {\bibfnamefont {T.}~\bibnamefont {Prosen}},\ }\href@noop {}
  {\  (\bibinfo {year} {2018})},\ \Eprint
  {http://arxiv.org/abs/arXiv:1805.00931} {arXiv:1805.00931} \BibitemShut
  {NoStop}%
\bibitem [{Note1()}]{Note1}%
  \BibitemOpen
  \bibinfo {note} {The restriction to a Gaussian-distributed disorder is not
  fundamental and the techniques developed here are easily generalised to any
  \protect \emph {divisible} disorder distribution.}\BibitemShut {Stop}%
\bibitem [{\citenamefont {Cotler}\ \emph {et~al.}(2017)\citenamefont {Cotler},
  \citenamefont {{Hunter-Jones}}, \citenamefont {Liu},\ and\ \citenamefont
  {Yoshida}}]{cotlerChaosComplexityRandom2017a}%
  \BibitemOpen
  \bibfield  {author} {\bibinfo {author} {\bibfnamefont {J.}~\bibnamefont
  {Cotler}}, \bibinfo {author} {\bibfnamefont {N.}~\bibnamefont
  {{Hunter-Jones}}}, \bibinfo {author} {\bibfnamefont {J.}~\bibnamefont {Liu}},
  \ and\ \bibinfo {author} {\bibfnamefont {B.}~\bibnamefont {Yoshida}},\
  }\href@noop {} {\bibfield  {journal} {\bibinfo  {journal} {J. High Energy
  Phys.}\ }\textbf {\bibinfo {volume} {2017}},\ \bibinfo {pages} {48} (\bibinfo
  {year} {2017})}\BibitemShut {NoStop}%
\bibitem [{Note2()}]{Note2}%
  \BibitemOpen
  \bibinfo {note} {For example, we can calculate $\rho (t)$ from $S_2(t) =
  \protect \mathbb {E}_{\protect \mathbf {x}}[ e^{it(H\otimes \protect \mathbb
  {I} - \protect \mathbb {I}\otimes H^T)}]$.}\BibitemShut {Stop}%
\bibitem [{Note3()}]{Note3}%
  \BibitemOpen
  \bibinfo {note} {The elementary derivation via Gaussian integrals of the main
  representation is carried out in the supplementary material.}\BibitemShut
  {Stop}%
\bibitem [{\citenamefont
  {\O{}ksendal}(2003)}]{oksendalStochasticDifferentialEquations2003}%
  \BibitemOpen
  \bibfield  {author} {\bibinfo {author} {\bibfnamefont {B.}~\bibnamefont
  {\O{}ksendal}},\ }\href@noop {} {\emph {\bibinfo {title} {Stochastic
  Differential Equations}}},\ \bibinfo {edition} {sixth}\ ed.,\ Universitext\
  (\bibinfo  {publisher} {{Springer-Verlag}},\ \bibinfo {address} {Berlin},\
  \bibinfo {year} {2003})\BibitemShut {NoStop}%
\bibitem [{\citenamefont
  {Gardiner}(1997)}]{gardinerHandbookStochasticMethods1997}%
  \BibitemOpen
  \bibfield  {author} {\bibinfo {author} {\bibfnamefont {C.~W.}\ \bibnamefont
  {Gardiner}},\ }\href@noop {} {\emph {\bibinfo {title} {Handbook of
  {{Stochastic Methods}}: For {{Physics}}, {{Chemistry}} and the {{Natural
  Sciences}}}}},\ \bibinfo {edition} {2nd}\ ed.,\ \bibinfo {series} {Springer
  Series in Synergetics}\ No.~\bibinfo {number} {13}\ (\bibinfo  {publisher}
  {{Springer-Verlag}},\ \bibinfo {address} {Berlin},\ \bibinfo {year}
  {1997})\BibitemShut {NoStop}%
\end{thebibliography}
\end{document}